\documentclass[twocolumn,amsmath,amssymb,10pt,superscriptaddress,a4paper,letterpaper,fleqn]{revtex4-1}
 \usepackage{amsmath}
 \usepackage{amssymb}
 \usepackage{epsfig}
 \usepackage{graphicx}
 \usepackage{dcolumn}
 \usepackage{array}
 \usepackage{bm}
                                        
 \usepackage{longtable}
 \usepackage{multirow}
 \usepackage{float}
                                      
 \usepackage{afterpage}
 \usepackage{color}
 \usepackage{footnote}
 \setlongtables
 \usepackage[breaklinks=true,linkbordercolor={1 1 1}]{hyperref}
 
\begin{document}
 \setcounter{page}{1}
 
 
 \title{Revealing Fine Structure in the Antineutrino Spectra From a Nuclear Reactor}

 \author{A.~A.~Sonzogni}
 \email{sonzogni@bnl.gov}                           
 \affiliation{National Nuclear Data Center, Bldg. 817, Brookhaven National Laboratory, Upton, NY 11973-5000, USA}

\author{M.~Nino}
\affiliation{Department of Physics and Astronomy, Hofstra University, Hempstead, NY 11549, USA}

 \author{E.~A.~McCutchan}
 \affiliation{National Nuclear Data Center, Bldg. 817, Brookhaven National Laboratory, Upton, NY 11973-5000, USA}

 \date{\today}
 
 \begin{abstract}
 {
We calculate the Inverse Beta Decay (IBD) antineutrino spectrum generated by nuclear reactors using the summation method  to
understand deviations from the smooth Huber-Mueller model due to the decay of individual fission products,
showing that plotting the ratio of two adjacent spectra points can effectively reveal these deviations.
We obtained that for binning energies of 0.1 MeV or lower, abrupt changes in the spectra  due
to the jagged nature of the individual antineutrino spectra could be observed for highly precise experiments.   
Surprisingly, our calculations also reveal  a peak-like feature in the adjacent points ratio plot at 4.5 MeV 
even with a 0.25 MeV binning interval, 
which we find is present in the IBD  spectrum published by Daya Bay in 2016.
We show that this 4.5 MeV feature is caused by the contributions of just four fission products, $^{95}$Y, $^{98,101}$Nb and $^{102}$Tc.
This would be the first evidence of the decay of a few fission products in the IBD antineutrino spectrum from a
nuclear reactor.
 }
 \end{abstract}
 \maketitle


Since the discovery of radioactivity in 1895~\cite{becquerel96}, scientists have been able to accurately characterize 
the $\gamma$, e$^-$, e$^+$, neutron, proton and $\alpha$  spectra emitted by nuclei during their radioactive decay.  
However, measuring the antineutrino spectra from the decay of individual nuclei has eluded experimental efforts even today as antineutrinos only interact with matter through the weak interaction.
The closest direct measurements of neutrino spectra following beta-decay are the  neutrino oscillation measurements of the SNO experiment~\cite{sno} which measured neutrinos following solar $^{8}$B decay. 
However, the energy distribution of antineutrinos from a single nucleus following $\beta$-minus decay has not been reported to the best of our knowledge. 
There has been, however, tremendous success in detecting antineutrinos from nuclear reactors, starting with the pioneering work of Cowan and Reines~\cite{cowan56}.  
Most recently, the Daya Bay~\cite{dayabay16}, Double Chooz~\cite{doublechooz}, and RENO~\cite{reno16} 
collaborations have measured reactor's antineutrino spectra with unprecedented statistics by placing large antineutrino detectors in the vicinity of commercial reactors.  
In these experiments, antineutrinos are produced  by the approximately 800 fission products,
and are later detected  through the use of the Inverse Beta Decay (IBD) reaction, $\overline{\nu}_e+p \rightarrow n+e^+$.  Since the IBD cross section increases steadily for
energies above its 1.8 MeV threshold, and as the antineutrino spectrum generated by a reactor decreases with increasing energy, the resulting IBD  spectrum has
a bell shape with the maximum around 3.5-4.0 MeV.

The focus in the interpretation and understanding of these  reactor experiments has been through comparing the integral and overall shape of the measured spectrum to various predictions.   
Currently, the antineutrino spectra from 
Huber~\cite{huber11} for  $^{235}$U,   $^{239}$Pu and  $^{241}$Pu, obtained from converting measured aggregate electron spectra~\cite{ill82,ill85,ill89},
and that of Mueller {\it et al.}~\cite{mueller11} for $^{238}$U derived from nuclear databases, are thought to be the best predictions.
Comparisons of measured IBD spectra with the Huber-Mueller model have led to the observation of an overall deficit in the number of measured antineutrinos along with a spectra distortion~\cite{mention11}.  These findings then spurred a flurry of investigations into the source of these disagreements~\cite{hayes14,hayes15,huber17,dayabay17,sonzogni17,mention17}.

To study the possible existence of sterile neutrinos, as well as to gauge antineutrinos' potential to monitor nuclear reactors~\cite{christensen14},
a new generation of experiments at very short baselines~\cite{prospect,nucifer,neos17} have begun or will soon begin to collect data near a reactor core.  
In particular, the NEOS collaboration~\cite{neos17}  recently published their IBD spectrum with a binning interval of 0.1 MeV, that is, a considerable improvement over the
0.25 MeV value that has been the standard so far.  
The presence of deviations in the NEOS IBD spectrum from the Huber-Mueller predictions near the maximum was the initial motivation for this work,
where we explore the possibility of observing  signatures of the individual fission products amidst the overall spectrum,
which in vague analogy with other radiation types, we will refer to as ``fine structure".
In an attempt to view the trees out of the forest, we present a novel approach to analyzing antineutrino spectra which involves taking ratios of adjacent 
energy bins.   
We then apply this technique to the highest resolution data available from the Daya Bay collaboration and find that even with a 0.25 MeV binning,  discontinuities in the antineutrino spectra are evident and moreover, are in agreement with calculations which consider the decay of all fission fragments produced in the reactor.  These discontinuities can be attributed to just a few nuclei, suggesting that indeed 
the individual signatures of the fission fragments can be unraveled from the whole.

Electrons and antineutrinos are produced in a nuclear reactor following 
the  $\beta-$ decay of neutron rich fission products, whose spectra can be calculated using the summation method~\cite{vogel81} as
\begin{equation}
  S(E) =  \sum_i \sum_j f_i \times CFY_{ij} \times S_j(E),
  \label{summ}
\end{equation}                   
\noindent where $f_i$ are the effective fission fractions for the four main fuel components,  $^{235}$U,  $^{238}$U,  $^{239}$Pu and  $^{241}$Pu; 
$CFY_{ij}$ are the cumulative fission yields from the neutron induced fission on these fissile nuclides; 
and  the spectra $S_j(E)$ are calculated as
\begin{equation}
  S_j(E) = \sum I_{jk} \times S_{jk}(E),
\end{equation}                   
\noindent with $I_{jk}$ the $\beta-$ decay intensity from the jth $\beta$-minus decaying level in the network to the kth level in the daughter nucleus, 
and $S_{jk}(E)$ are the corresponding nuclear level to nuclear level spectra, 
which for electrons are given by
\begin{equation}
\begin{split}
  S(E)=& N \times W \times (W^2-1)^{1/2} \times (W-W_0)^2  \times \\& F(Z,W) \times C(Z,W),
\end{split}
\label{l2lespec}
\end{equation}                     
\noindent where N is a constant so that $S(E)$ is normalized to unity;
$W$ is the relativistic kinetic energy, 
$W=E/m_ec^2 +1$, and $W_0$=$Q/m_ec^2 +1$, with $Q$ the total decay energy available also known as
the end-point energy; 
$F(Z,W)$ is the Fermi function and $Z$ is the number of protons in the daughter nucleus; lastly,  the $C(Z,W)$ term 
contains the corrections due to angular momentum and parity changes,  finite size, screening, radiative and weak magnetism.

\begin{figure}[t] 
\includegraphics[width=0.95\columnwidth, trim=20mm 0mm 15mm 5mm, clip=true]{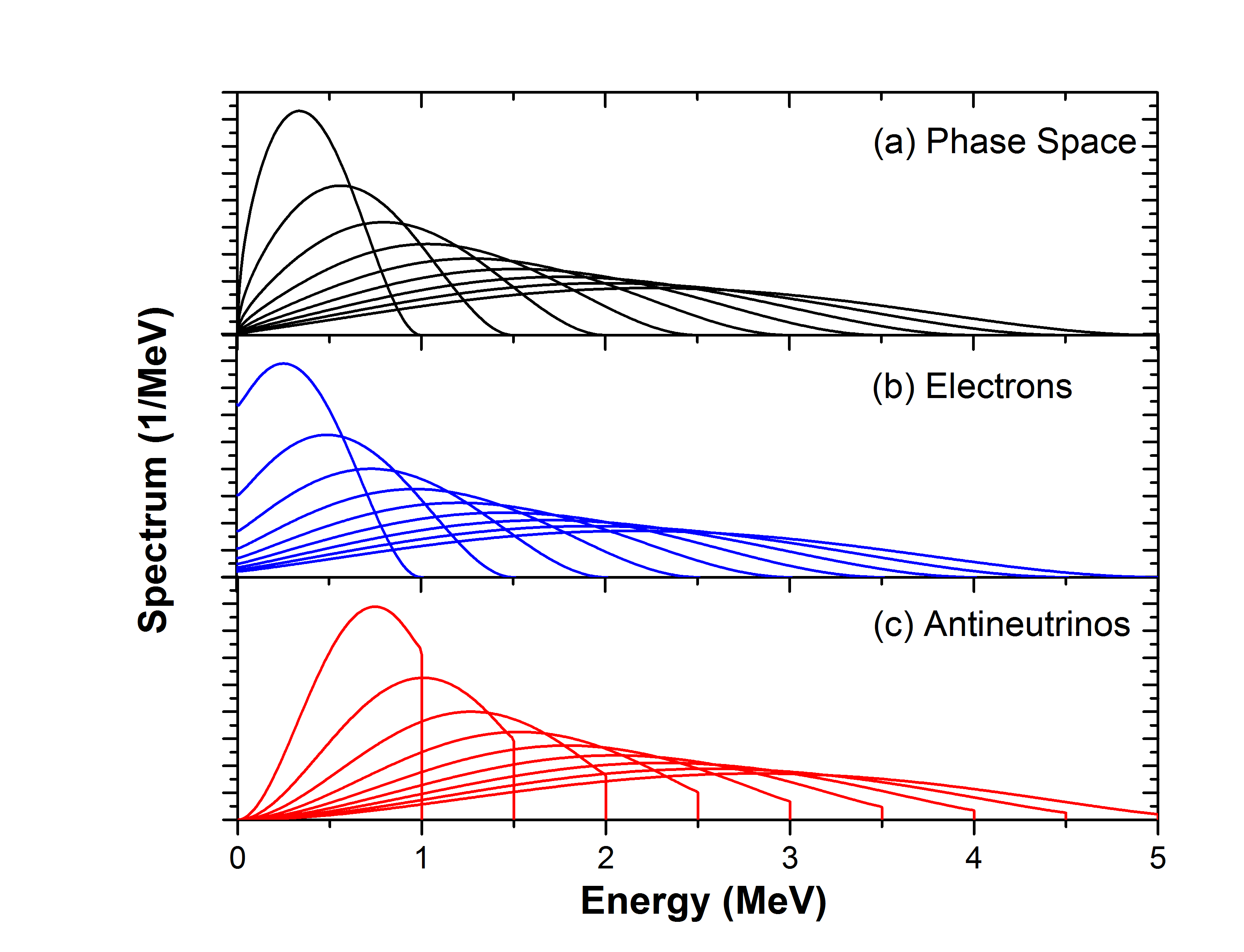}
\caption{
(a) Energy distribution using only the phase space term, 
(b) energy distribution for electrons including all the correction terms  (c) as the middle panel but for antineutrinos.
}
\label{f.spectra}
\end{figure}   

The $W \times (W^2-1)^{1/2} \times (W-W_0)^2 $ product in Eq. (\ref{l2lespec}) is colloquially known as the ``phase space" term as it is customarily inferred using statistical classical mechanics 
arguments.   In the absence of Coulomb effects, the electron and antineutrino spectra would be given by this term, resulting in identical energy distributions, as seen in Fig.~\ref{f.spectra}~(a).
The presence of the Coulomb field generated by the protons in the nucleus gives rise to the Fermi function term, F(Z,W), which slows down the electrons. 
Conservation of energy then dictates a corresponding boost to the antineutrino energy. 
The effect for
electrons can be seen in Fig.~\ref{f.spectra}~(b), while the bottom panel shows the antineutrino spectra.   
Because of the Fermi function, the level-to-level antineutrino spectra exhibit an abrupt cutoff at
the end-point energy.  As a consequence, while a sum of electron spectra will have a smooth quality, a sum of antineutrino spectra may have a rugged, serrated nature.
Fig.~\ref{f.spectra} also reveals that the cut-off in the antineutrino spectrum is more severe for lower-energy antineutrinos, being particularly pronounced for end point energies between 1 and 3 MeV, 
and with increasing end point energy the serrated portion of the spectrum quickly diminishes.  

\begin{figure}[t] 
\includegraphics[width=0.95\columnwidth, trim=55mm 10mm 55mm 20mm, clip=true]{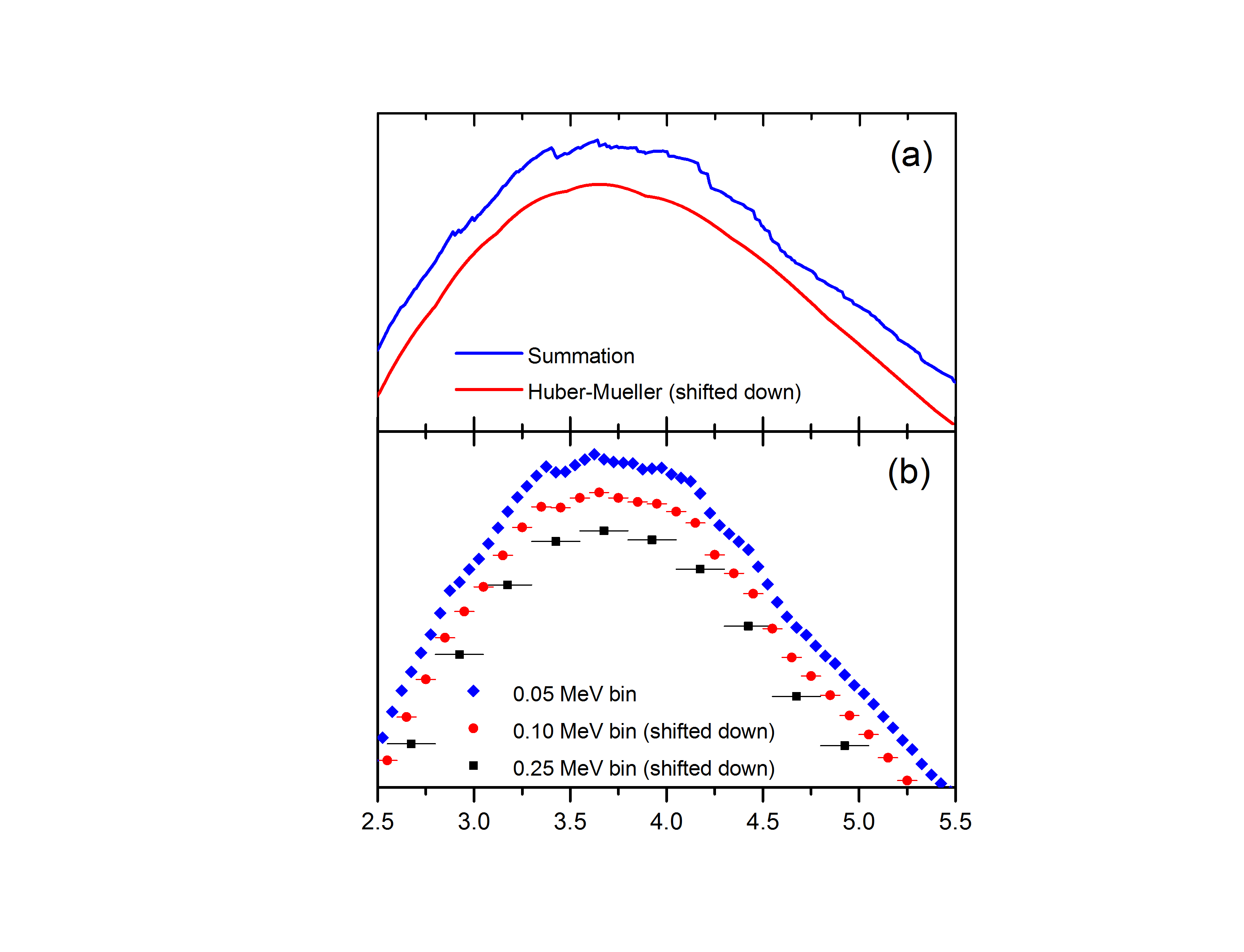}
\caption{
(a) Calculated Daya Bay IBD antineutrino spectra using the summation method (top blue line) 
and the Huber-Mueller antineutrino spectra (lower red line), shifted down to highlight the jaggedness of the former versus the smoothness of the latter.
(b) IBD antineutrino spectra from summation calculations binned in 0.05, 0.10 and 0.25 MeV intervals.   For contrast reasons, the last two curves are shifted down;
additionally, experimental resolution effects were not included.  
}
\label{f.bin}
\end{figure}

Producing a neutron-rich source  of a single isotope, strong enough to observe the intriguing features  in Fig. 1~(c), would be experimentally extremely challenging. The question is, within the 800 fission fragments  
(with 10 to 100's of $\beta$ branches) produced in a nuclear reactor, can the spectra from individual nuclei be disentangled. 
We explore this possibility by calculating the antineutrino spectra for the Daya Bay experiment with the
summation method using:
(a) Daya Bay fission fractions~\cite{dayabay16};
(b) JEFF-3.1 fission yields~\cite{jeff31} and updated  ENDF/B-VII.1 decay data~\cite{e71} 
as described in Refs.~\cite{sonzogni15,sonzogni16};
(c) $F(Z,W)$  and $C(Z,W)$ obtained as described by Huber~\cite{huber11};
(d) IBD cross sections from Ref.~\cite{xs}. 
The resulting IBD antineutrino spectrum is shown in Fig.~\ref{f.bin}(a), 
plotted together for contrast with the corresponding Huber-Mueller spectra (shifted down).   The energy range was restricted
to 2.5 to 5.5 MeV, where statistical uncertainties will be sufficiently low to allow the observation of fine structure.

Hints of fine structure begin to materialize in the summation calculation starting around 2.8 MeV; 
the most obvious is a sharp cutoff just below 3.5 MeV, which corresponds to the sharp cutoffs seen in Fig.~1(c). 
There is also another type of structure that spans several hundreds of keV, such as a shoulder around 4.2 to 4.5 MeV, which would correspond to the contribution of a single fission product or a group of them that have similar end point energies.
The effect of binning in our summation calculations, not including experimental resolution effects, is explored in Fig.~\ref{f.bin}~(b), where with 0.25 MeV binning intervals, 
the standard one so far for Daya Bay, deviations from a smooth curve would be hard to discern by eyesight, while with 0.10 MeV bins, as available with the recent NEOS data~\cite{neos17}, fine structure begins to manifest, and would become considerably clearer with a 0.05 keV binning.  
A similar analysis as a function of the detector resolution was presented in Ref.~\cite{dwyer15}.


\begin{figure}[t] 
\includegraphics[width=0.95\columnwidth, trim=15mm 10mm 25mm 17mm, clip=true]{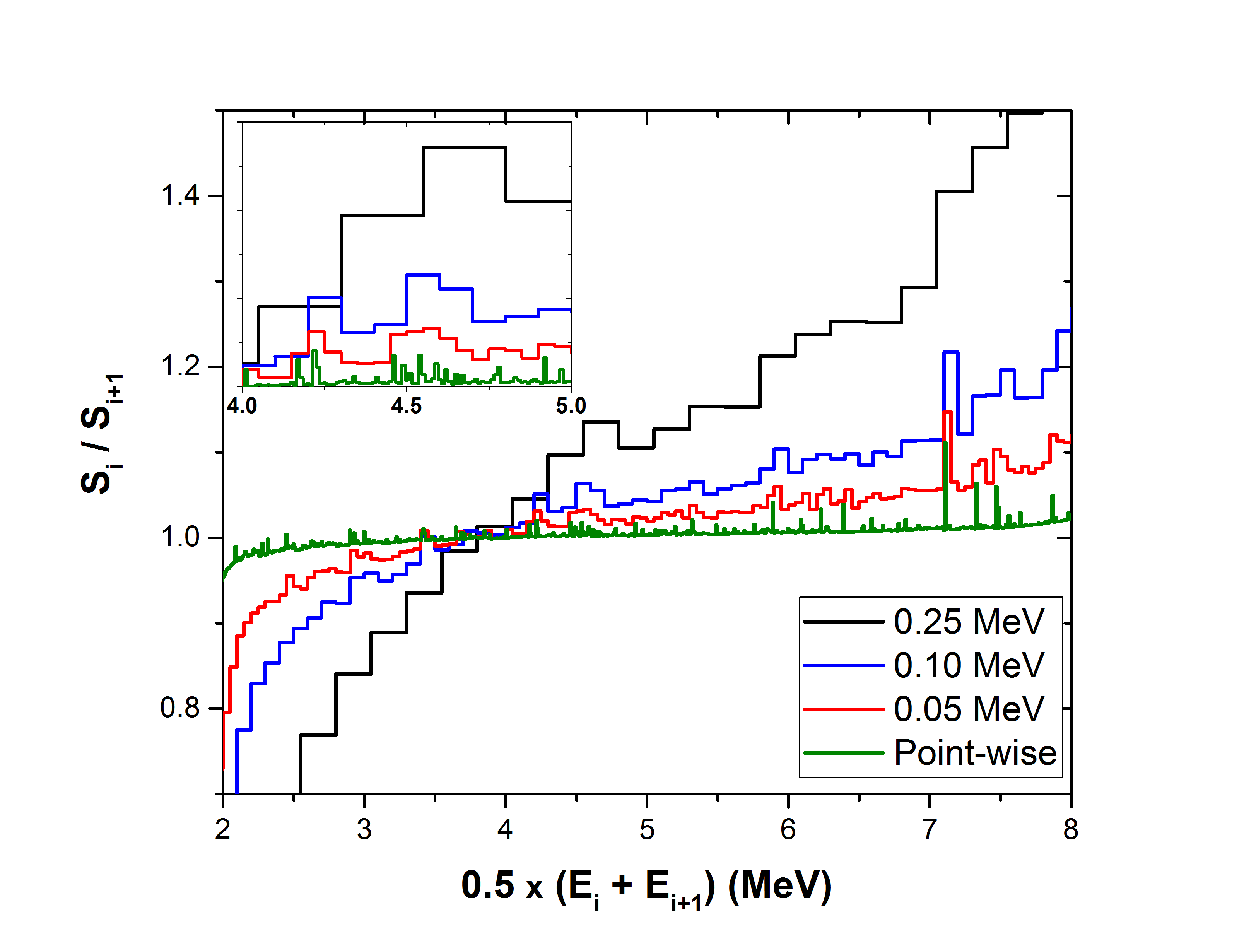}
\caption{
Ratio of two consecutive summation IBD spectrum points under different binning conditions.
A zoom in the  4 to 5 MeV region can be seen in the top-left corner inset.
}
\label{f.ratio}
\end{figure}   

\begin{figure}[t] 
\includegraphics[width=0.95\columnwidth, trim=5mm 5mm 25mm 5mm, clip=true]{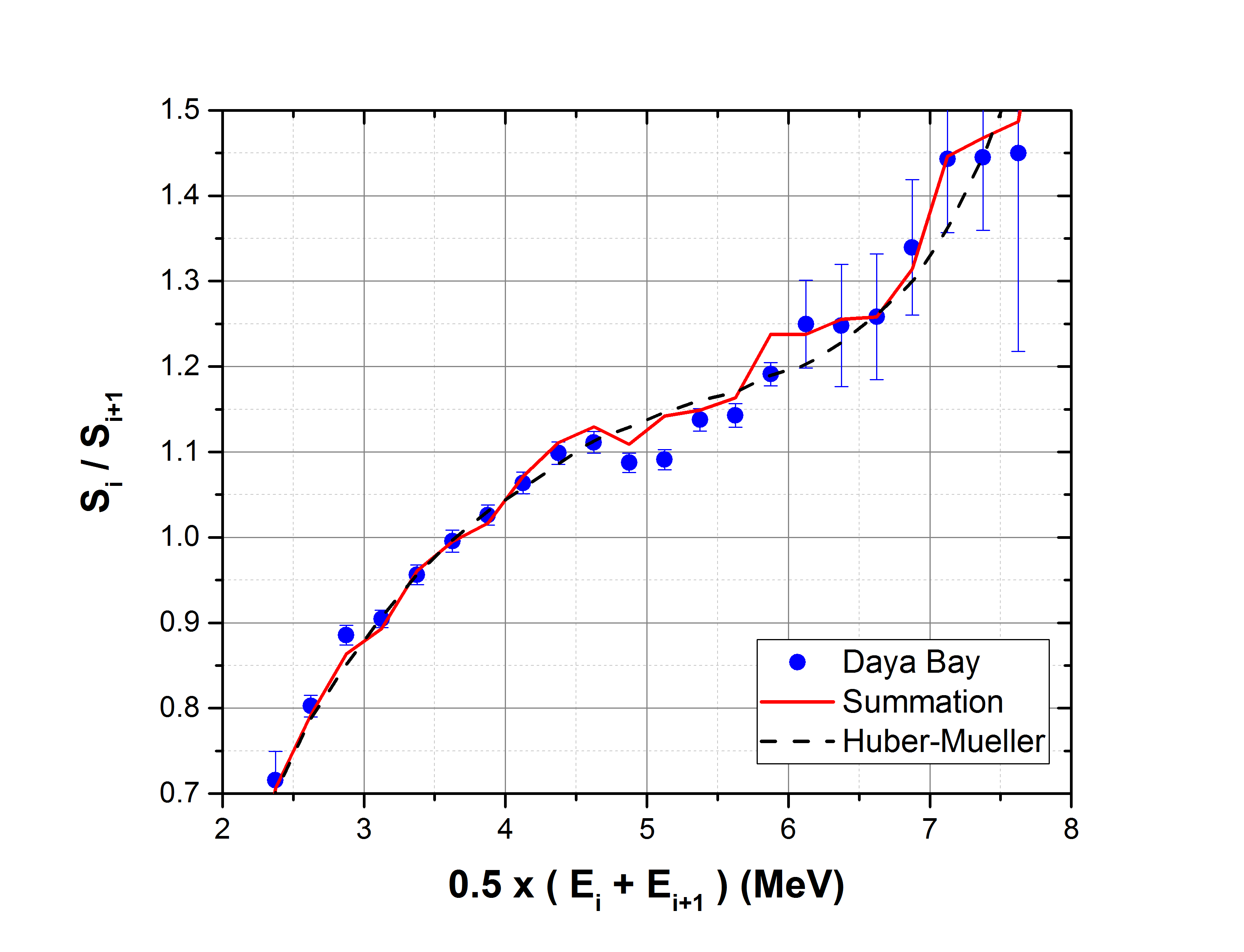}
\caption{
Ratio of two consecutive IBD spectrum points from the Daya Bay experiment (symbols), Huber-Mueller model (dashed black line) and
summation method (full red line). 
For clarity's sake, the calculated points are connected with a straight line, instead of a step function as in Fig.~\ref{f.ratio}. 
}
\label{f.dbratio}
\end{figure}

While there are definite features in the spectra shown in Fig.~\ref{f.bin},  they are difficult to interpret quantitatively.  
To better elucidate fine structure features, we consider the ratio between two adjacent points
\begin{equation}
  R_i =  S_i / S_{i+1},
\end{equation}
\noindent as a function of the average bin interval $0.5 \times ( E_i+E_{i+1} )$,
which are plotted  in Fig.~\ref{f.ratio} for a summation calculation IBD spectrum,
under different binning scenarios.   
For a point-wise calculation, the sudden drop in the spectrum due to the abrupt end of a relevant fission product antineutrino spectrum 
would manifest as a peak, for instance that of $^{96}$Y at 7.1 MeV, whose signature can also be seen with  0.05 and 0.1 MeV binning.
Intuitively, it may be thought that for a 0.25 MeV binning, a smooth curve would be obtained.   
However,  
a peak-like feature around 4.5 MeV is visible, which as the inset in the 
top-left corner reveals, it is caused by a number of strong transitions that coincidentally have similar end-point energies.
Inspired by this observation,  the Daya Bay $R_i$ values are plotted in Fig.~\ref{f.dbratio}, where the uncertainties on $R_i$ were calculated using a first order Taylor expansion
\begin{equation}
\begin{split}
  \Delta^2 R_i =  & S^{-2}_{i+1} \times \sigma_{i,i}   + S^2_i \times S^{-4}_{i+1} \times \sigma_{i+1,i+1} \\& - 2 \times S_i \times S^{-3}_{i+1} \times \sigma_{i,i+1}  ,
\end{split}
\end{equation}                   
\noindent with $\sigma$ the covariance matrix as given by the Daya Bay Collaboration~\cite{dayabay16}. 
The need of the covariance matrix is significant here since as the adjacent $S_i$ points are positively correlated,
that is  $\sigma_{i+1,i+1} > 0$, the $ \Delta R_i$ values are smaller than those obtained assuming no correlation, 
which allows for a positive identification of the structures in the $R_i$ plot.

While the summation calculation $\chi^2$ per point in Fig.~\ref{f.dbratio} is only marginally smaller than the corresponding Huber-Mueller $\chi^2$, the summation
calculation shape is remarkably more similar to the experimental one.  This
demonstrates the necessity to improve
the summation method, which
despite being less precise than the conversion method due to deficiencies in fission yield and decay data~\cite{fallot12}, is absolutely needed to
fully understand the features of a reactor antineutrino spectrum.

\begin{figure}[t] 
\includegraphics[width=0.95\columnwidth, trim=22mm 2mm 22mm 6mm, clip=true]{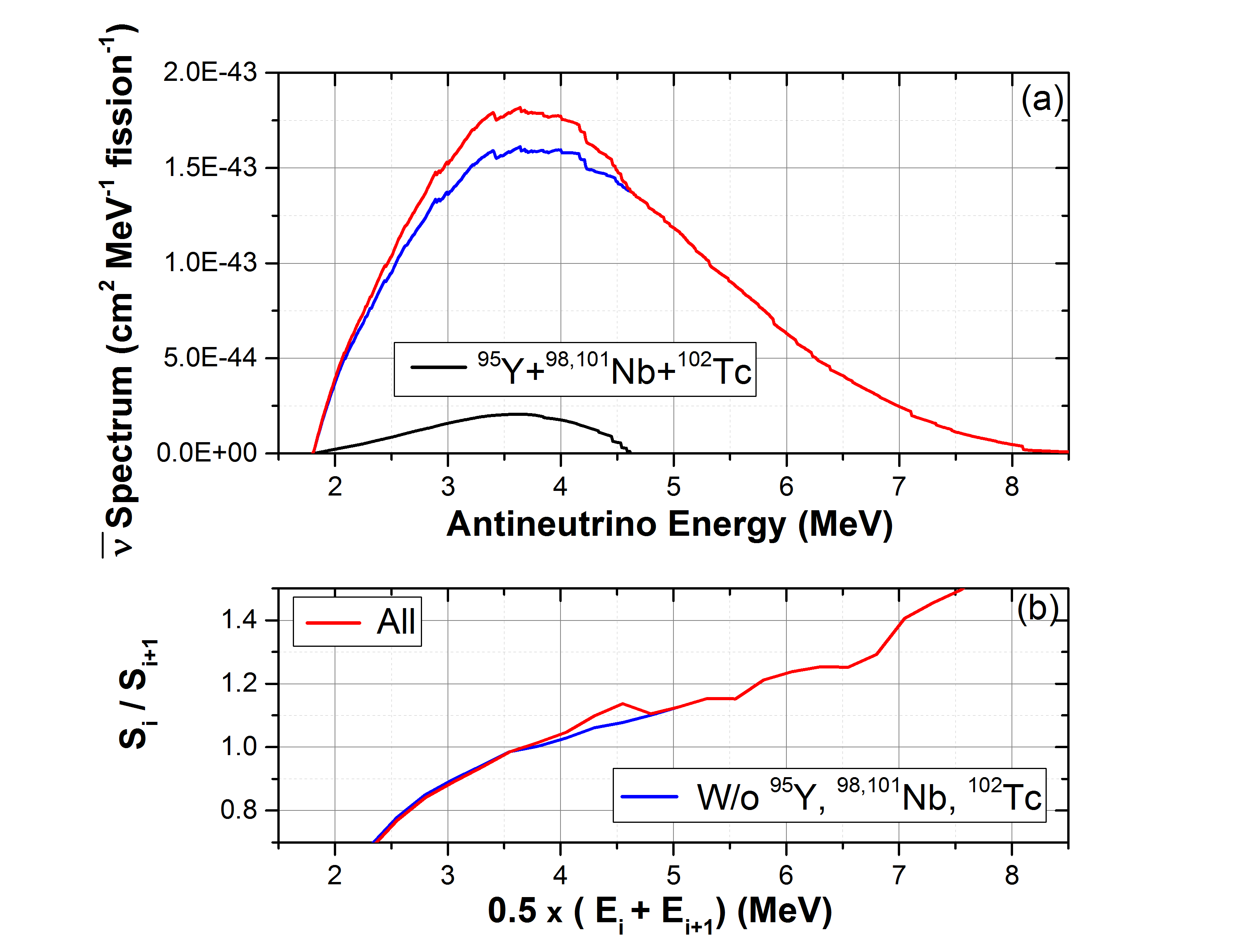}
\caption{
 (a) Calculated Daya Bay IBD antineutrino spectra from all the fission products (red line), the $^{95}$Y, $^{98,101}$Nb, and $^{102}$Tc contribution (black line),
and the difference (blue line).
(b) Corresponding ratio of two adjacent points with a 0.25 MeV binning.
}
\label{f.hl}
\end{figure}

The next question is if can we attribute the 4.5 MeV peak-like feature to individual nuclei. To answer this question,
we searched for the most relevant individual IBD spectra with large $R$ values in the 4 to 5 MeV region.
We find that the feature at 4.5 MeV is caused by just four nuclides,
$^{95}$Y, $^{98,101}$Nb, and $^{102}$Tc, which have in common 
large cumulative fission yields and antineutrino spectra 
dominated by strong transitions with end-point energy near 4.5 MeV. 
To understand why the number is so low, we need to remember that due to the double-humped nature of
independent fission yield distributions, there is a relatively small number of nuclides with significant effective cumulative fission yields, $CFY_{eff,i}=\sum f_k CFY_{ki}$.
For instance, for Daya Bay fission fractions, the largest effective CFY for products contributing to the IBD spectrum is that of $^{134}$I with 
$CFY_{eff,i}$=0.073,
and while the number of nuclides with effective CFYs larger than
0.01 is about 115, it drops to about 30 for effective CFYs larger than 0.05.    
This number is further reduced when we require these nuclides to have large values of $I_{ki}$ with end-point energies in the 4 to 5 MeV region.
Fig.~\ref{f.hl}~(a) shows the total IBD spectrum, the one generated by the four nuclides in question,
and the difference.   
These four nuclides account for about 6\% of the total IBD antineutrino yield, and about 9.6\% of the IBD antineutrino yield in the 1.8 to 4.5 MeV region.
Fig.~\ref{f.hl}~(b) shows a plot of $R_i$ values with a 0.25 MeV binning with and without the contribution of $^{95}$Y, $^{98,101}$Nb, and $^{102}$Tc,
which clearly shows that the feature at 4.5 MeV is basically caused by these four nuclides.

As several prior works have investigated the need of reliable decay and fission yield data~\cite{fallot12,sonzogni15,rasco16} to accurately calculate 
antineutrino spectra,
we now assess the quality of the  $^{95}$Y, $^{98,101}$Nb, and $^{102}$Tc data
to determine if the 4.5 MeV structure 
in the summation calculations is a solid prediction.  
The effective JEFF CFYs for these nuclides, 
with their relative uncertainty in parenthesis, are  0.058 (0.9\%), 0.057 (3.4\%), 0.054 (1.9\%), and  0.050 (3.7\%) respectively.  
The ENDF/B-VII.1 effective cumulative fission yields are similar, within 2\% or less from the JEFF values, with the exception of $^{102}$Tc, whose 
independent fission yield is considerably smaller than its cumulative as it is mainly produced in the decay of $^{102}$Mo.   
When the ENDF/B yields were obtained, it was assumed that the isomer would take most of the $\beta$-decay intensity, which as we know today it is not the case.
When this correction is applied, both values of cumulative fission yields agree very well.
A look at the decay data used in our calculations reveals that for these four nuclides, the $\beta$ intensity pattern is dominated by 
a strong ground state (GS) to ground state transition with end point energies near 4.5 MeV.
For instance, GS to GS transitions intensities and Q-values are: 
64$\pm$1.7\% and 4.45 MeV for $^{95}$Y~\cite{a95},
 57$\pm$7\% and 4.59 MeV for $^{98}$Nb~\cite{a98},
40$\pm$13\% and 4.55 MeV for $^{101}$Nb~\cite{a101},
92.9$\pm$0.6\% and 4.53 MeV for $^{102}$Tc~\cite{a102,jordan13}.
We conclude that the nuclear data for these four nuclides are fairly reliable due to the relative
closeness to the valley of stability.

\begin{figure}[t] 
\includegraphics[width=0.95\columnwidth, trim=0mm 0mm 15mm 0mm, clip=true]{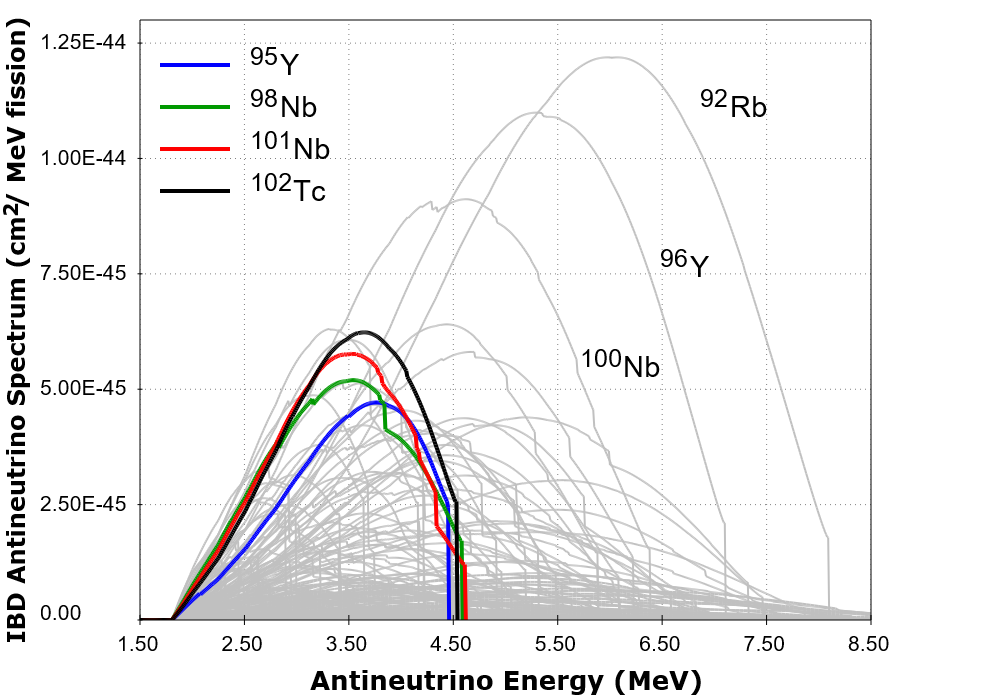}
\caption{
Calculated IBD antineutrino spectra from all the fission products, highlighting
 the $^{95}$Y, $^{98,101}$Nb, and $^{102}$Tc ones.
}
\label{f.spaghetti}
\end{figure}   

Further insights  can be obtained by studying Fig.~\ref{f.spaghetti}, where the IBD antineutrino spectra for all the fission products
are plotted, highlighting in particular the $^{95}$Y, $^{98,101}$Nb  and $^{102}$Tc ones, whose sum spectrum, due to the similarity of the end-point energies,
 effectively looks like that of a single strongly produced fission product.  As a comparison,
the three large spectra observed on the right
side of the plot are, in descending order, those from
$^{92}$Rb, $^{96}$Y and $^{100}$Nb, which contribute about 6.7\%, 5.3\% and 4.1\% to the total IBD antineutrino yield, respectively.   
Despite their sizable contribution, observing the fine structure caused
by their antineutrino spectra sharp cutoff would be considerably 
more difficult as the relative importance of the sharp cutoff diminishes and the IBD antineutrino spectrum is considerably smaller at energies close to their end-point energies. 

In summary, we have shown that as the IBD antineutrino spectrum from a nuclear reactor is the sum of about 800 fission products, 
there will be deviations from a smooth shape due to 
(a) the contribution of a strongly populated fission product, 
(b) sharp cutoffs in the individual antineutrino spectra,
and (c) the contribution from a small number of fission products with similar end-point energy, effectively mimicking the first case.
We developed a novel, yet simple way of numerically revealing these contributions, by plotting the ratio of adjacent points. 
We conclude that with a binning interval of 0.1 MeV or less, the observation of sharp cutoffs from the individual spectra could be observed.
Remarkably, even with a binning of 0.25 MeV, we are able to detect a peak-like feature in the ratio plot, which we can attribute to the decay of 
$^{95}$Y, $^{98,101}$Nb and $^{102}$Tc.
This exercise clearly shows the need for highly reliable fission and decay data to fully understand all the features in the IBD antineutrino spectrum from 
a nuclear reactor.

\begin{acknowledgements} 
Work at Brookhaven National Laboratory was sponsored by the Office of Nuclear Physics, Office of Science of the U.S.  
Department of Energy under Contract No. DE-AC02-98CH10886, and by the
DOE Office of Science, Office of Workforce Development
for Teachers and Scientists (WDTS), under the Science
Undergraduate Laboratory Internships Program (SULI).
This work was partially supported under the U.S.
Department of Energy FIRE Topical Collaboration in Nuclear
Theory.

We are grateful to the Daya Bay collaboration for publishing their experimental points, including the covariance matrix, which have made possible  the analysis presented in this article.

\end{acknowledgements}


\begin{thebibliography}{10}

\bibitem{becquerel96} M.~H.~Becquerel, C. R. Physique {\bf 122}, 420 (1896).

\bibitem{sno} Q.~R.~Ahmad {\it et al.}, Phys. Rev. Lett. {\bf 87}, 071301 (2001).

\bibitem{cowan56} C.~L.~Cowan, Jr., F.~Reines, F.~B.~Harrison, H.~W.~Kruse, A.~D.~McGuire, Science, {\bf 24}, Number 3212, 103  (1956). 
 
\bibitem{dayabay16} F.~P. An  {\it et al.}, Phys. Rev. Lett. {\bf 116},  061801 (2016).

\bibitem{doublechooz} Y. Abe {\it et al.},  Phys. Rev. Lett. {\bf 108}, 131801 (2012).  

\bibitem{reno16} J.~H.~Choi {\it et al.}, Phys. Rev. Lett. {\bf 116}, 211801 (2016).

\bibitem{huber11} P. Huber,  Phys. Rev. C {\bf 84} 024617 (2011).

\bibitem{ill82} F. von Feilitzsch, A.~A.~Hahn, and K.~Schreckenbach, Phys. Lett. B {\bf 118}, 162 (1982).

\bibitem{ill85} K. Schreckenbach {\it et al.}, Phys. Lett. B {\bf 160}, 325 (1985).

\bibitem{ill89} A.~A. Hahn {\it et al.,} Phys. Lett. B {\bf 218}, 365 (1989).
 
\bibitem{mueller11} T.~A. Mueller {\it et al.,} Phys. Rev. C {\bf 83}, 054615 (2011). 

\bibitem{mention11} G. Mention {\it et al.,}  Phys. Rev. D {\bf 83}, 073006 (2011). 

\bibitem{hayes14} A.~C.~Hayes {\it et al.,} Phys. Rev. Lett. {\bf 112}, 202501 (2014).

\bibitem{hayes15} A.~C. Hayes {\it et al.,} Phys. Rev. D {\bf 92}, 033015 (2015).

\bibitem{huber17} P. Huber, Phys. Rev. Lett. {\bf 118}, 042502 (2017).

\bibitem{dayabay17} F.~P. An  {\it et al.}, Phys. Rev. Let. {\bf 118}, 251801 ( 2017).

\bibitem{sonzogni17} A.~A.~Sonzogni, E.~A.~McCutchan, and A.~C.~Hayes, Phys. Rev. Lett.  {\bf 119}, 112501 (2017).

\bibitem{mention17} G.~Mention, M.~Vivier, J.~Gaffiot, T.~Laserre, A.~Letourneau, T.~Materna, Phys. Lett. B {\bf 773}, 307 (2017).

\bibitem{christensen14}  E. Christensen, P. Huber, P. Jaffke, and T.E. Shea, Phys. Rev. Lett. {\bf 113}, 042503 (2014).

\bibitem{prospect} J. Ashenfelter {\it et al.,} Nucl. Instrum. Methods Phys. Res., Sect. A {\bf 806}, 401 (2016).

\bibitem{nucifer} G. Boireau {\it et al.},  Phys. Rev. {\bf D} 93, 112006 (2016).

\bibitem{neos17} Y.~Ko {\it et al.}, Phys. Rev. Lett. {\bf 118}, 121802 (2017).

\bibitem{vogel81} P. Vogel {\it et al.}, Phys. Rev. C {\bf 24}, 1543 (1981).

\bibitem{jeff31} M.~A. Kellett, O. Bersillon, and R.W. Mills, {\sc JEFF report 20}, OECD, ISBN 978-92-64-99087-6 (2009).  

\bibitem{e71} M.~B. Chadwick {\it et al.}, Nucl. Data Sheets {\bf 112}, 2887 (2011).    

\bibitem{sonzogni15} A.~A.~Sonzogni, T.~D.~Johnson, and E.~A.~McCutchan, Phys. Rev. C {\bf 91}, 011301(R) (2015).

\bibitem{sonzogni16} A.~A.~Sonzogni, E.~A.~McCutchan, T.~D.~Johnson, and P.~Dimitriou, Phys. Rev. Lett.  {\bf 116}, 132502 (2016).

\bibitem{xs} A. Strumia and F. Vissani, Phys. Lett. B {\bf 564}, 42 (2003). 

\bibitem{dwyer15} D.~A.~Dwyer and T.~J.~Langford, Phys. Rev. Lett {\bf 114}, 012502 (2015).

\bibitem{fallot12} M.~Fallot {\it et al.},  Phys. Rev. Lett. {\bf 109}, 202504 (2012).

 \bibitem{rasco16} B.~C.~Rasco {\it et al.}, Phys. Rev. Lett. {\bf 117}, 092501 (2016). 

\bibitem{a95} S. K. Basu, G. Mukherjee, and A. A. Sonzogni, Nucl. Data Sheets {\bf 111}, 2555 (2010).

\bibitem{a98}  B. Singh, and Z. Hu, Nucl. Data Sheets {\bf 98}, 335 (2003).

\bibitem{a101} J. Blachot, ENSDF database, www.nndc.bnl.gov/ensdf.

\bibitem{a102} D. De Frene, Nucl. Data Sheets {\bf 110}, 1745 (2009).

\bibitem{jordan13} D. Jordan {\it et al.}, Phys.Rev. C {\bf 87}, 044318 (2013)


\end{thebibliography}
\end{document}